\documentclass[12pt]{iopart}

\usepackage[figuresright]{rotating}
\usepackage{floatflt}
\usepackage{graphicx}
\usepackage{dcolumn}
\usepackage{bm}

\newcommand{\AmS}{{\protect\the\textfont2
  A\kern-.1667em\lower.5ex\hbox{M}\kern-.125emS}}

\def\Journal#1#2#3#4{{#1} {\bf #2}, #3 (#4)}

\def\PLB{{ Phys. Lett.}  B}
\def\PRL{ Phys. Rev. Lett.}

\def\PRC{{ Phys. Rev.} C}
\def\PRD{{ Phys. Rev.} D}

\def\JPG{{ J. Phys.} G}

\def\IPE{{\em Int. J. Mod. Phys.} E}

\begin{document}

\title[Overview of charm production at RHIC]{Overview of charm production at RHIC}

\author{Yifei Zhang}
\address{Dept. of Modern Physics, University of Science and Technology of China, Hefei, Anhui, China, 230026}
\ead{yfzhang3@mail.ustc.edu.cn}

\begin{abstract}

We present an overview of the recent abundant measurements for
charm production at RHIC. The significant information of charm
cross sections in different collision system at 200 GeV and
charmed hadron freeze-out and flow properties extracted from these
measurements are presented. The heavy flavor energy loss in the
medium and heavy flavor related azimuthal correlations in heavy
ion collisions are also discussed.

\end{abstract}
\vspace{-0.5cm}
\pacs{25.75.Dw, 13.20.Fc, 13.25.Ft, 24.85.+p}

\section{Introduction}
\vspace{-0.35cm}

Charm quarks are a unique tool to probe the hot-dense matter
created in relativistic heavy-ion collisions at RHIC. Charm quarks
are believed to be produced only in the early stages and their
production rate is reliably calculable by perturbative
QCD~\cite{lin,cacciari}. Studies of the binary collision
($N_{bin}$) scaling of the total charm cross section can be used
to test theoretical calculations and determine if charm is indeed
a good probe with well-defined initial states. Measurements of
charm production at low $p_{T}$, in particular radial and elliptic
flow, probe the QCD medium and are thus sensitive to bulk medium
properties like density and the drag constant or viscosity. And
charm flow properties may help understand the light flavor
thermalization~\cite{teaney}. Due to their large mass ($\simeq$1.3
GeV/$c^2$), charm quarks are predicted to lose less energy than
light quarks by gluon radiation in the medium~\cite{dead}. Measure
heavy quark energy loss via its semileptonic decay electrons may
provide us information on the interactions of heavy quarks with
the hot dense matter produced in nuclear collisions at RHIC. The
strong energy loss of light hadrons modified the di-hadron
azimuthal correlation functions~\cite{dihcorr}. Measurement of the
heavy flavor electron-hadron correlation may help look insight the
mechanism of heavy quark interactions with the hot dense medium.
Bottom quark is much heavier than charm quark, their energy loss
and flow behavior may be very different. The separation of the
bottom and charm contribution in the non-photonic electron
measurement is vital to clearly understand the charm and bottom
quark production and their interactions with the medium.

In this paper, instead of going through the details for all the
measurements or analyses, we prefer the discussion on those hot
physics about charm cross section, heavy quark flow, energy loss
and heavy flavor related correlations from recent abundant
observations at RHIC.

\section{D-mesons and leptons from heavy flavor decays}

STAR experiment has measured open charm via hadronic channel
($D^0\rightarrow K\pi$, B.R.=3.83\%) at low $p_T$ ($<\sim3$
GeV/$c$) in $d$+Au~\cite{dAuCharm}, minbias Cu+Cu~\cite{AlexQM08}
and minbias Au+Au~\cite{STARAuAuCharm} collisions at 200 GeV. Good
signals (${}_\sim^{>}4\sigma$) are observed in the $K\pi$
invariant mass distributions after the combinatorial background
subtraction using event-mixing method. Another hadronic channel
($D^0\rightarrow K\pi\pi^0$, B.R.=14.1\%) to reconstruct open
charm was measured by PHENIX experiment in $p+p$ collisions at 200
GeV~\cite{PHENIXD}. In this measurement, the $\pi^0$ was
identified by EMCal trigger via $\pi^0\rightarrow\gamma\gamma$
decay, and $\sim3\sigma$ signal was seen in the $p_T$ range of
5-15 GeV/$c$.

Due to the difficulty to reconstruct D-meson hadronic decay vertex
using current detectors, both STAR and PHENIX have measured open
charm indirectly via its semileptonic decays to electrons or
muons. STAR measured non-photonic electrons using
TPC+TOF~\cite{dAuCharm,STARAuAuCharm} and TPC+EMC~\cite{STAREMCe},
these two results are consistent with each other but are
systematically higher than PHENIX results, which were measured
using RICH and EMC in a lower material
environment~\cite{PHENIXeloss,PHENIXppe,PHENIXAuAue}. But the
nuclear modification factors ($R_{AA}$) of electrons from heavy
quark decays are consistent between STAR and PHENIX. STAR TOF has
the capability to measure single muon at very low $p_T$ range
(0.17-0.21 GeV/$c$) at mid-rapidity, which constrain 90\% of the
charm production cross section~\cite{ffcharm,STARAuAuCharm}.
PHENIX measured muons at high $p_T$ ($>2$ GeV/$c$) using forward
muon detector at rapidity $\langle y\rangle=1.65$~\cite{PHENIXmu}.
In addition, the di-electron from heavy flavor decays in $p+p$
collisions at 200 GeV has been measured by PHENIX experiment using
a cocktail method. At mid-rapidity, the charm and bottom cross
section are derived from the comparison between the $e^+e^-$
invariant mass distributions from data and those from PYTHIA
simulations~\cite{PHENIXdie}.

\section{Charm production cross section}

\begin{figure}[htp]
\centering
\includegraphics[width=1.85in]{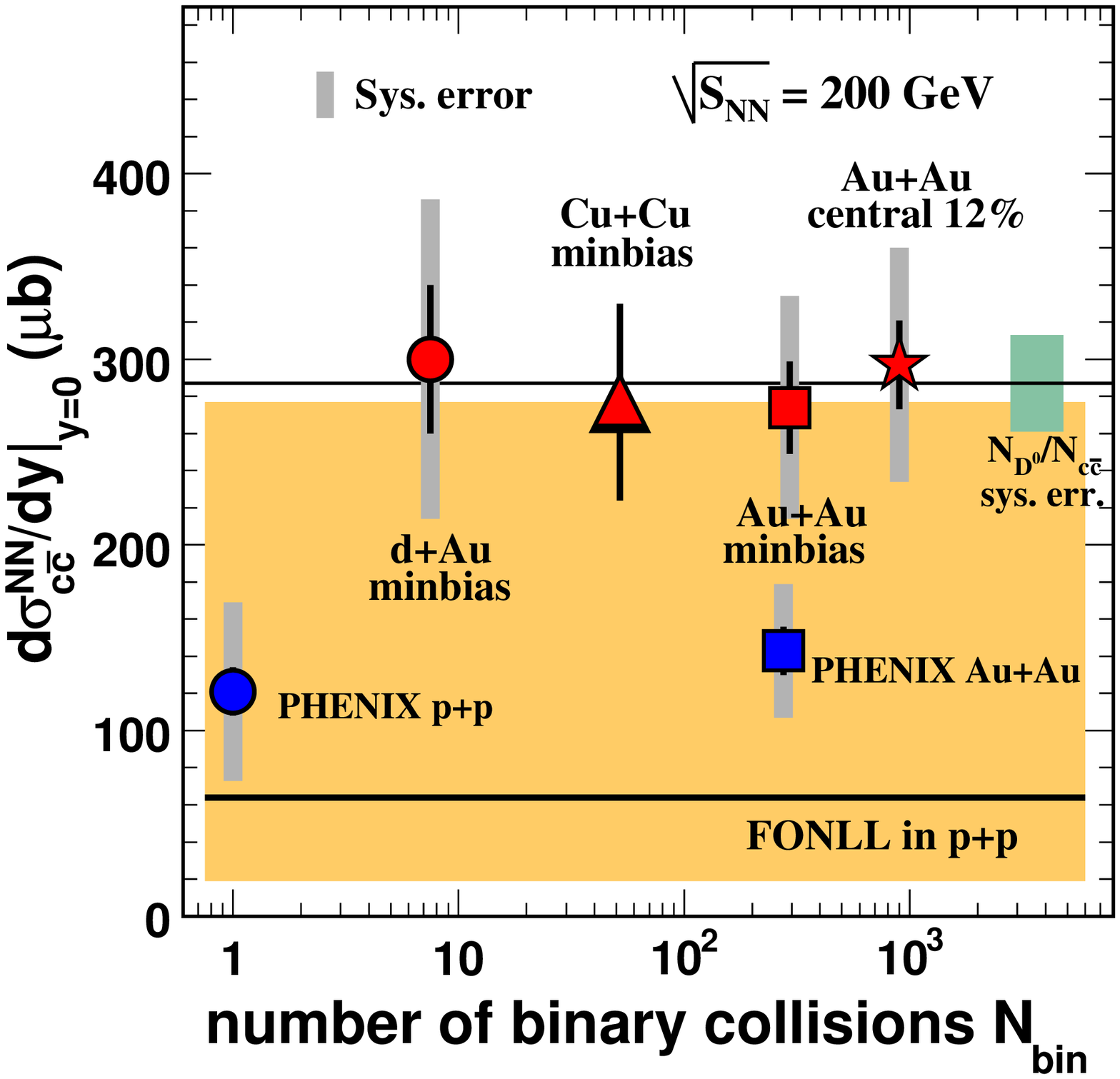}
\includegraphics[width=2.5in]{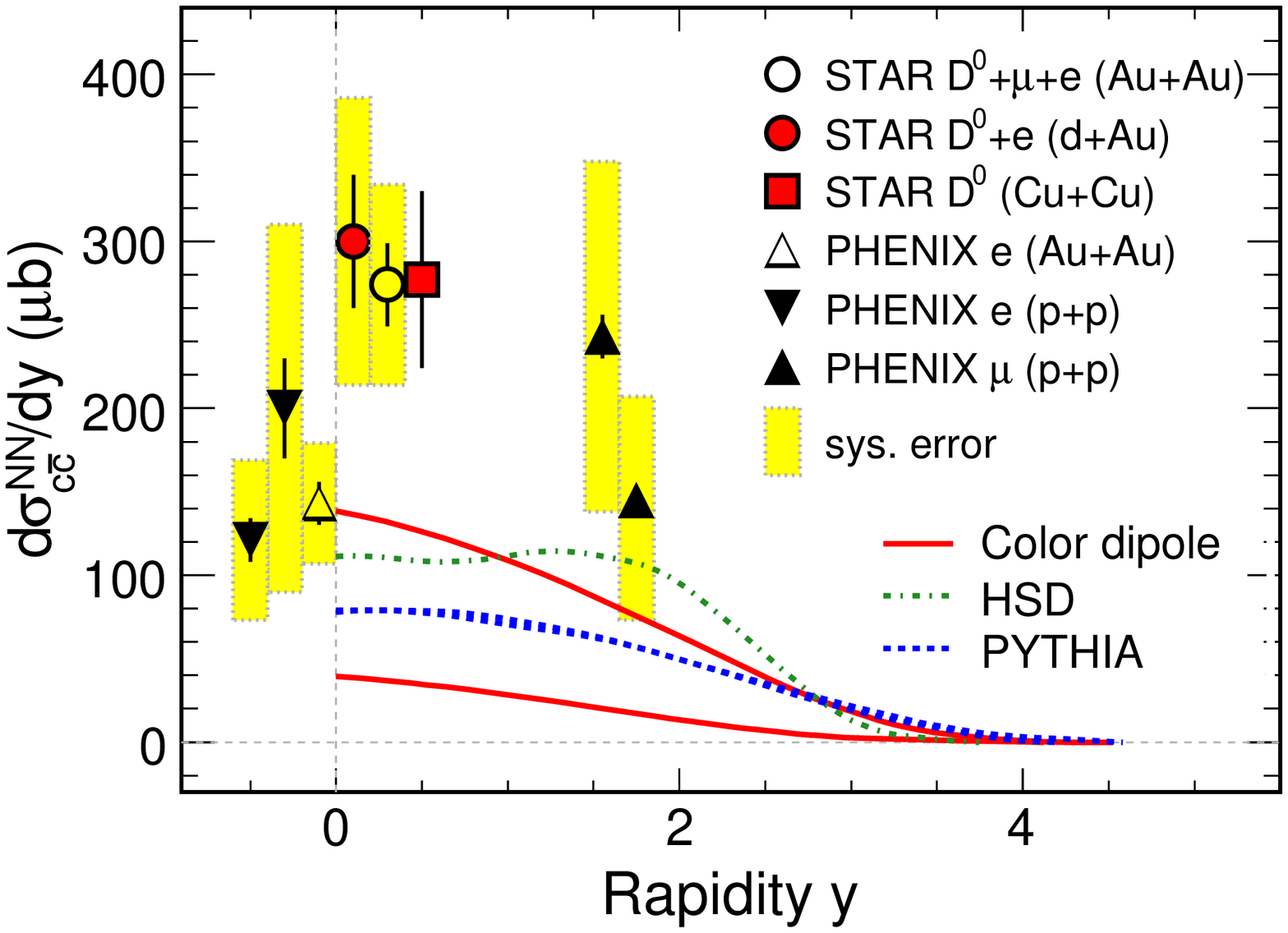}
\centering \caption{Panel (a):Mid-rapidity charm cross section per
nucleon-nucleon collision as a function of $N_{bin}$ in $d$+Au,
minbias and 0$-$12\% central Au+Au collisions. The solid line
indicates the average. FONLL prediction is shown as a band around
the central value (thick
line)~\cite{vogt}.}\vspace{-0.25cm}\label{fig:cxsec}
\end{figure}

Both STAR and PHENIX have measured charm production cross section
in several collision systems. Left panel of Fig.~\ref{fig:cxsec}
shows $d\sigma_{c\bar{c}}^{NN}/dy$ as a function of number of
binary collisions $N_{bin}$. The STAR previous result, the charm
production cross section at mid-rapidity in $d$+Au
collisions~\cite{dAuCharm} is shown as the red circle. The charm
cross section in Au+Au minbias collisions, derived by combining
three independent measurements of $D^0\rightarrow K\pi$ ($p_T<3$
GeV/$c$), muon ($0.17<p_T<0.21$ GeV/$c$) from charm decay and
non-photonic electron ($0.9<p_T<4$ GeV/$c$) from heavy flavor
decays, is shown as the red square. The result in Au+Au central
collisions (red star) is from combing the muon and electron
measurements~\cite{STARAuAuCharm}. The new result from Cu+Cu
minbias collisions is obtained from the $D^0$ ($p_T<3.3$ GeV/$c$)
measurement with statistics only (red triangle)~\cite{AlexQM08}.
The results from non-photonic electron measurements in 200 GeV
$p+p$ ($0.3\leq$$p_T$$\leq9.0$ GeV/$c$)~\cite{PHENIXppe} and Au+Au
($0.4\leq$$p_T$$\leq4.0$ GeV/$c$)~\cite{PHENIXAuAue} collisions at
PHENIX, are shown as the blue circle and the blue square,
respectively. The charm production cross section at mid-rapidity
scales with number of binary interactions both in STAR and PHENIX
experiments. This indicates that charm quarks are produced in the
early stage of relativistic heavy-ion collisions. The FONLL
calculation~\cite{vogt} shown as the band. Both the STAR and
PHENIX results are higher than the central value (thick line) of
the FONLL calculation, but the upper theory value reproduces the
experimental results. The central values of the cross sections
reported by PHENIX~\cite{PHENIXppe,PHENIXAuAue} are a factor of
about two smaller than STAR at all measured $p_T$~\cite{STAREMCe}.
The difference is approximately 1.5 times the combined
uncertainties, also shown in the right panel of
Fig.~\ref{fig:cxsec} at mid-rapidity. Right panel of
Fig.~\ref{fig:cxsec} shows the charm cross section as a function
of rapidity compared to theoretical calculations~\cite{RP03}, the
clear difference is seen between STAR and PHENIX results at
mid-rapidity with systematical errors dominated. PHENIX also
obtained the charm cross section from muon measurement at forward
rapidity ($\langle y\rangle=1.65$, $1.0\leq$$p_T$$\leq3.0$
GeV/$c$) in 200 GeV $p+p$ collisions, shown as the triangles. The
new result has smaller systematical error and consistent with
theory curves~\cite{PHENIXmu}.

\section{Flow and energy loss}

\begin{figure}[htp]
\centering
\includegraphics[width=1.7in]{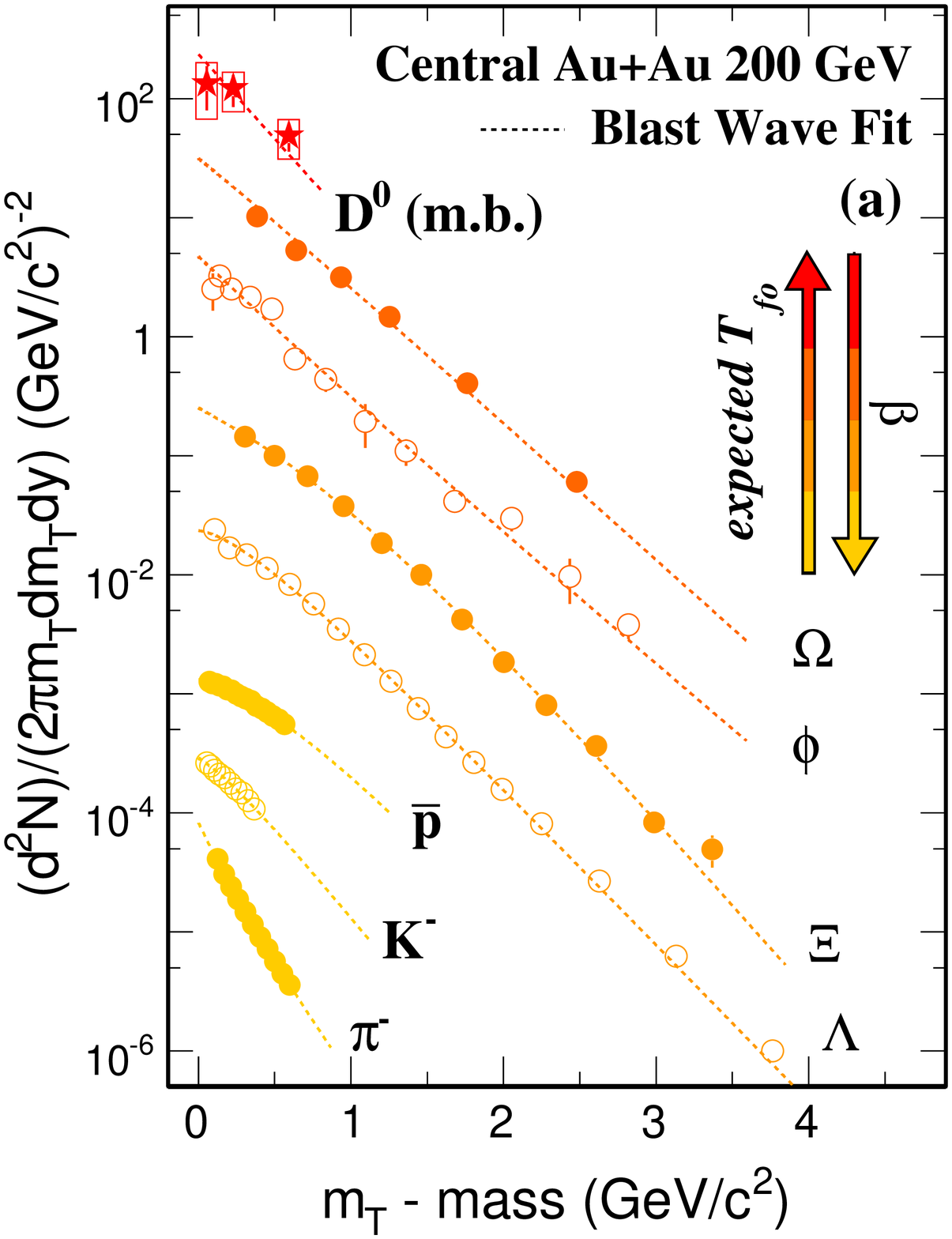}
\includegraphics[width=2.3in]{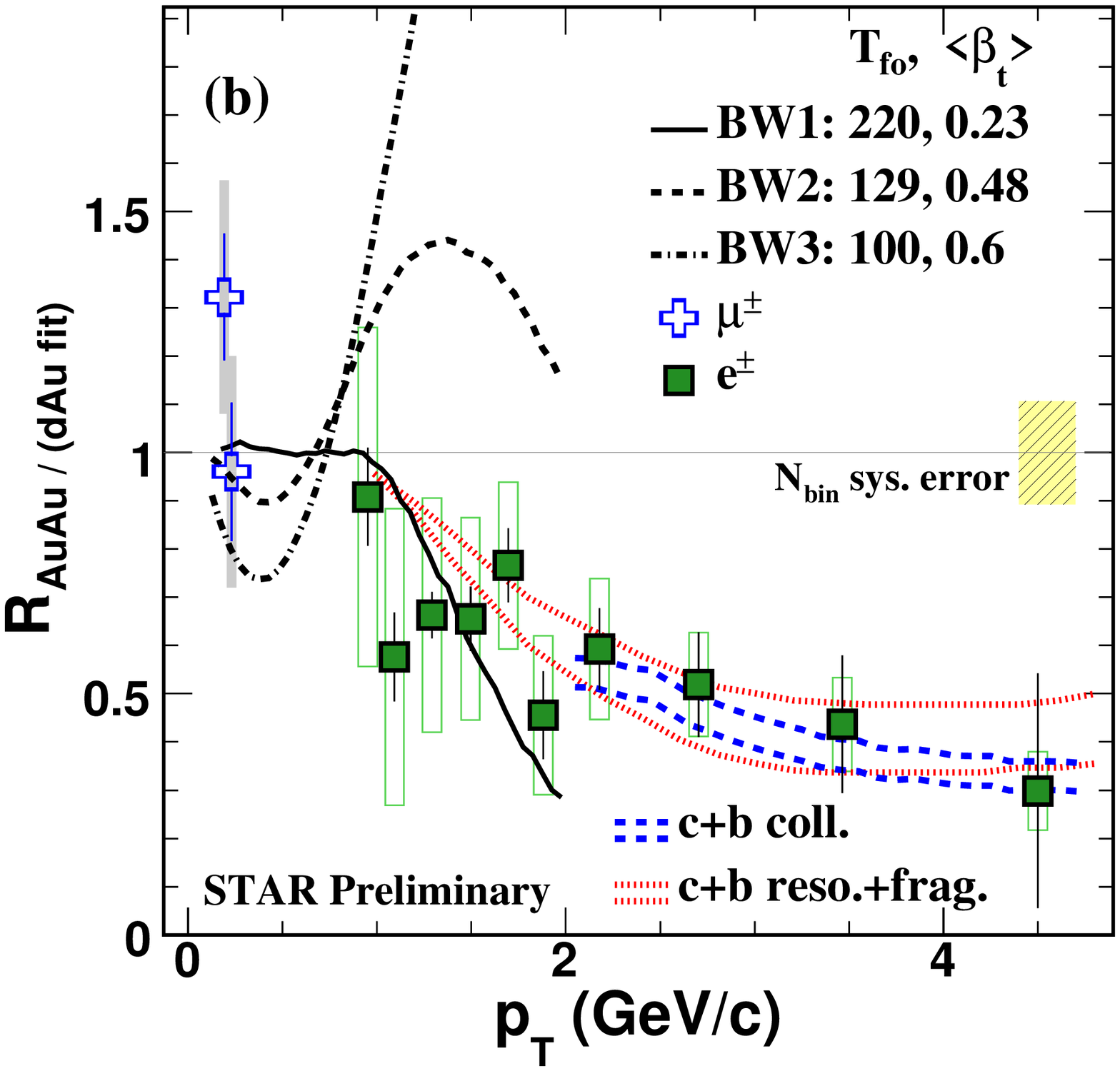}
\centering \caption{Panel (a): Hadron species dependent freeze-out
and flow parameters from blast wave fits to the hadron $m_T$
spectra. Panel (b): Nuclear modification factor ($R_{AuAu/dAu}$)
for 0$-$12\% Au+Au
collisions.}\vspace{-0.25cm}\label{fig:floweloss}
\end{figure}

In the hot dense medium created at Au+Au collisions, heavy quark
is consider as an intruder, put into the hot medium with
relatively very high density of light quarks. Due to their large
mass, such a heavy quark may acquire flow from the sufficient
interactions with the constituents of a dense medium in analog to
Brownian motion. Theoretical calculations have shown that
interactions between the surrounding partons in the medium and
heavy quarks could change the measurable
kinematics~\cite{teaney,HGR06}, and could boost the radial and
elliptic flow resulting in a different heavy quark $p_T$ spectrum
shape. Panel (a) of Fig.~\ref{fig:floweloss} shows the $m_T$
spectra for light hadrons ($\pi$, K ,p), $\Lambda$, $\Xi$ and
multi-strange hadrons ($\phi$, $\Omega$) in 200 GeV central Au+Au
collisions~\cite{thermalhadron,hyperon}, and charmed hadron
($D^0$) in 200 GeV minbias Au+Au collisions in
symbols~\cite{STARAuAuCharm}. Due to relatively heavy quark mass
and smaller hadronic scattering cross section, heavy flavor
hadrons are expected to freeze-out early and difficultly
participate in collective motion. Thus the larger freeze-out
temperature and smaller flow velocity are expected for heavy
flavor hadrons. The blast wave~\cite{SSH93} fit results (curves)
show significant dependence of the hadron species. From bottom
(light hadrons) to top (charmed hadron), the freeze-out
temperature is getting higher and the flow velocity becomes
smaller, shown as the arrows, which is consistent with our
expectations.

Panel (b) shows $R_{AuAu/dAu}$ for 0$-$12\% Au+Au collisions. To
study whether charmed hadrons have similar radial flow to light
hadrons, we have included curves for the expected nuclear
modification factor from a blast-wave model, using the freeze-out
parameters for light hadrons~\cite{thermalhadron} (BW3 in
Fig.~\ref{fig:floweloss} Panel (b)) and multi-strange
hadrons~\cite{hyperon} (BW2). The data and best blast-wave fit
(BW1) show large deviations from both these curves for $p_{T}>1$
GeV/$c$, which suggests that the charmed hadron freeze-out and
flow are different from light hadrons. We scanned the parameters
to a 2-dimensional $T_{fo}$, $\langle \beta_t \rangle$ space, the
results show little sensitivity to freeze-out temperature, but
disfavor large radial flow. These findings, together with the
observation of large charm elliptic flow~\cite{PHENIXeloss}, are
consistent with the recent prediction from
hydrodynamics~\cite{hirano}: elliptic flow is built up at partonic
stage, and radial flow dominantly comes from hadronic scattering
at later stage where charm may have already decoupled from the
system.

Since there is no direct charmed meson measurement at high $p_T$
currently, the $R_{AA}$ of non-photonic electrons from heavy
flavor decays was used to reveal heavy quark energy loss
indirectly. The strong suppression similar to light hadrons of the
non-photonic electrons $R_{AA}$ at high ${p_{T}}_{\sim}^{>}4$
GeV/$c$ has been observed in several
experiments~\cite{STAREMCe,PHENIXeloss}. In this case, STAR and
PHENIX are consistent with each other. Theoretical calculation
predicts that heavy quark lose less energy in the medium than
light quarks due to small gluon radiation angle~\cite{dead}. As
presented in Fig.~\ref{fig:floweloss} Panel (b), model
calculations of coalescence and fragmentation~\cite{RappRaa}
(double-dotted curves), and collisional dissociation of heavy
meson~\cite{Ivancoll} (double-dashed curves) describe the
experimental data.

\begin{figure}[htp]
\centering
\includegraphics[width=2.3in]{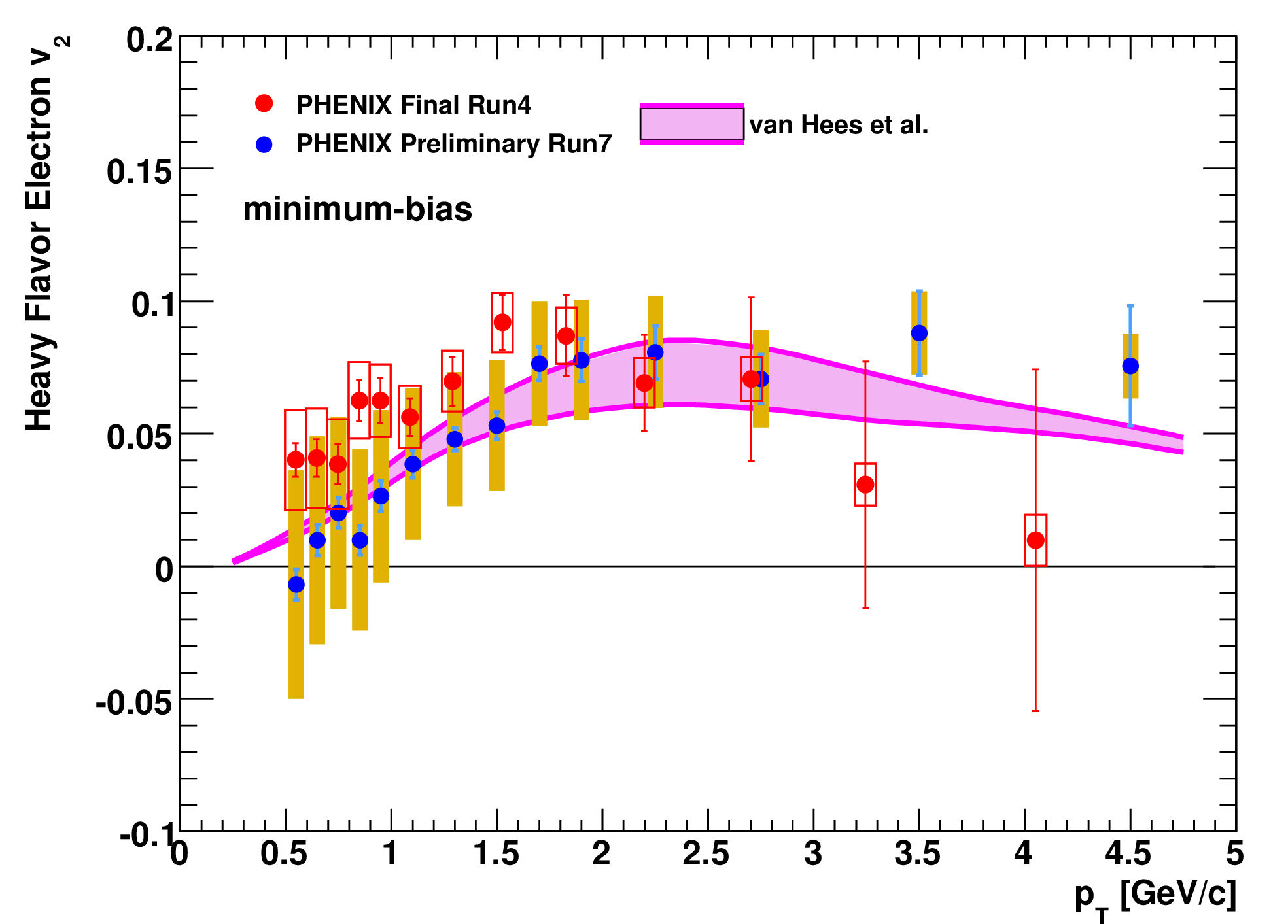}
\centering \caption{Non-photonic electron elliptic flow $v_2$ in
minbias Au+Au collisions at 200 GeV. Error bars are statistical
and the shaded boxes are systematical
errors.}\vspace{-0.25cm}\label{fig:ehcorr2}
\end{figure}

In the mean time, due to the strong angular correlation between
heavy flavor hadron and electron from its semileptonic decay at
high $p_T$, the non-photonic electron elliptic flow $v_2$ can be
used to measure heavy flavor hadron $v_2$. None zero $v_2$ of
electron from heavy flavor decays was observed at PHENIX. Fig.
shows the non-photonic electron $v_2$ measured from
run4~\cite{PHENIXeloss} and run7~\cite{DionQM08}. This may
indicate that the heavy quark strongly interact with the dense
medium at early stage of heavy ion collisions and the partonic
level collective motion has been observed at RHIC. The none zero
but smaller non-photonic electron $v_2$ is consistent with the
none zero but smaller radial flow velocity $\langle \beta_t
\rangle$ compared to light hadrons. Both the observations may
suggest that the light flavor thermalization at partonic stage in
the hot dense matter created in heavy ion collisions.

\section{Correlations}

The large amount of energy loss of high $p_T$ partons in the dense
medium created in central $A+A$ collisions was observed at
RHIC~\cite{lqeloss}. And their azimuthal correlations with low
$p_T$ hadrons are also modified by interacting with the medium,
showing a broad or even double-peak structure on the away-side
di-hadron correlation~\cite{dihcorr}. Since the similar level of
non-photonic electron energy loss as light hadrons was observed at
RHIC~\cite{STAREMCe,PHENIXeloss}, the study of the e-h azimuthal
correlation distribution could help us to understand the mechanism
of heavy quark energy loss in the dense medium and the
corresponding correlation pattern.

\begin{figure}[htp]
\centering
\includegraphics[width=2.2in]{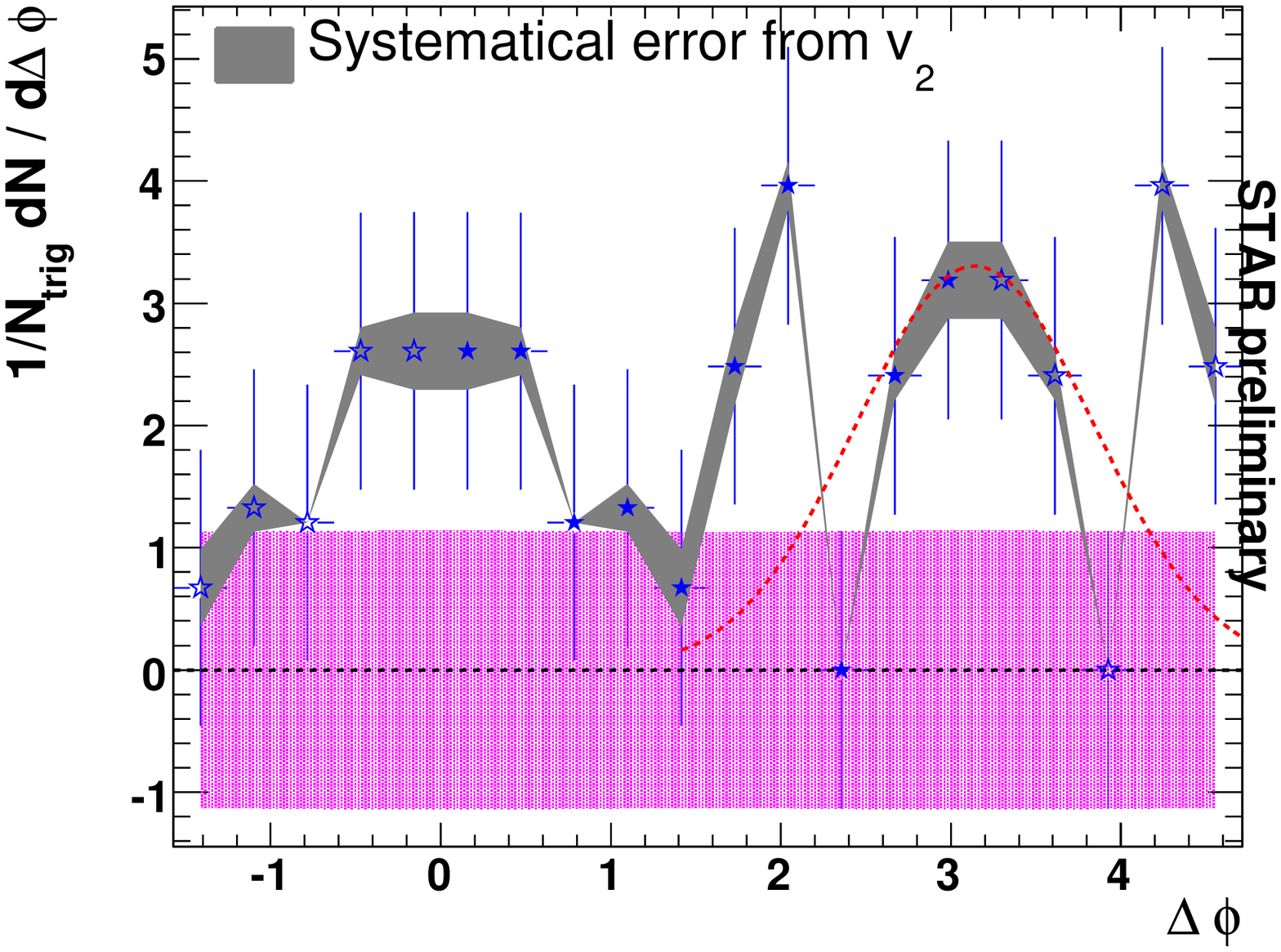}
\includegraphics[width=2.5in]{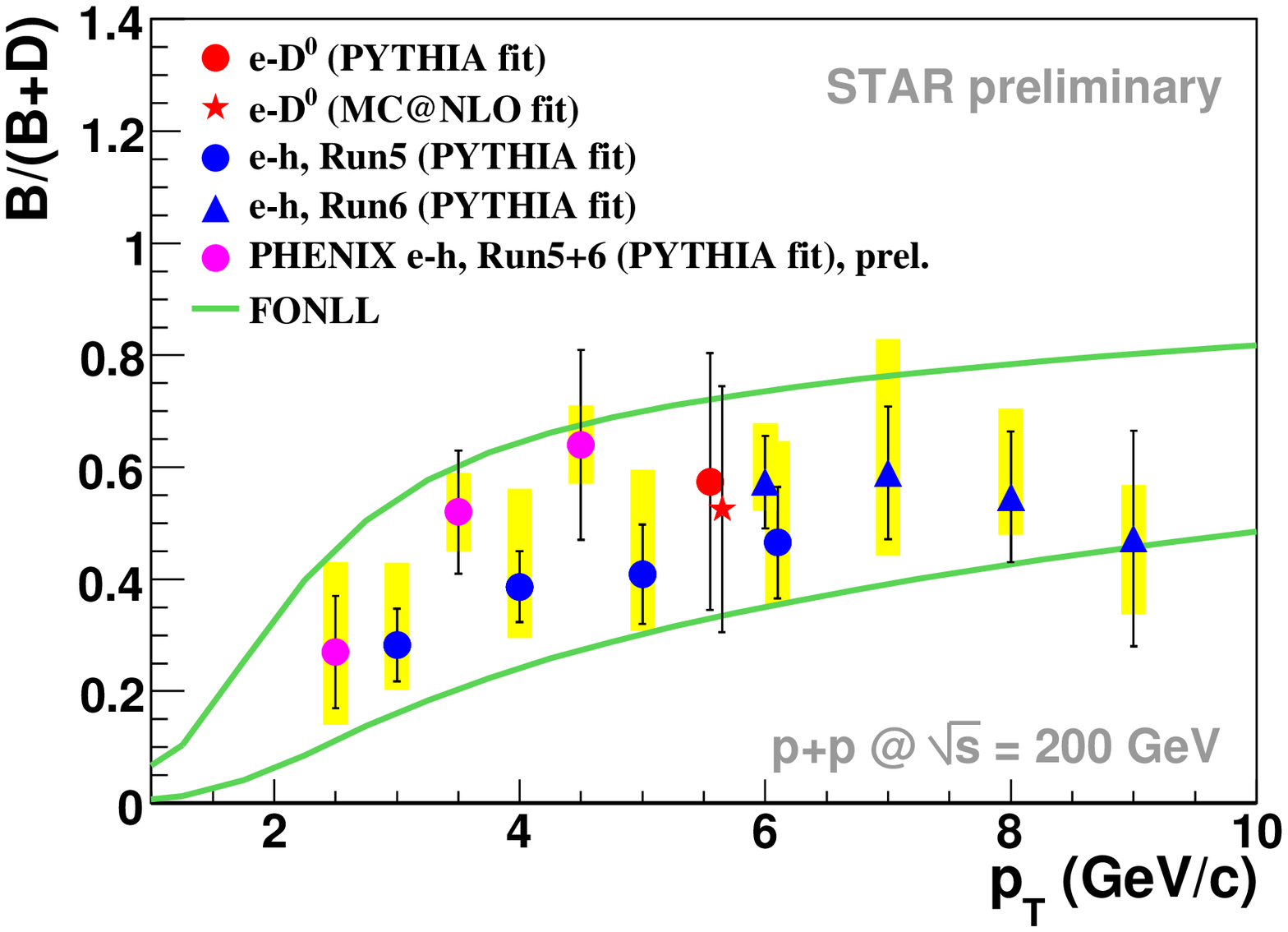}
\centering \caption{Left Panel: Non-photonic electron-hadron
correlation in Au+Au collisions at 200 GeV after the $v_2$
background subtraction. The dashed curve fitting to the data is
from PYTHIA expectations on the away side. The error bars are
statistical, and the error band around zero shows the systematical
uncertainty. Right Panel: Relative bottom contribution to the
total non-photonic electron yield derived from e-h and e-$D^0$
correlations.}\vspace{-0.25cm}\label{fig:ehcorr1}
\end{figure}

The azimuthal angular correlation of non-photonic electron and
charged hadrons has been measured at STAR~\cite{GangQM08}. Left
panel of Fig.~\ref{fig:ehcorr1} shows non-photonic e-h
correlations at 200 GeV Au+Au collisions. Here the $v_2$
background has been subtracted. Despite large statistical errors,
the azimuthal correlation distribution shows clear structure. On
the near side, there is one single peak representing the heavy
quark fragmentation, and possible interactions with the medium. On
the away side, instead of one peak around $\pi$ as in $p+p$
collisions, the correlation functions are modified to be a broad
even a double-peak structure, which is similar to the di-hadron
correlation in Au+Au~\cite{dihcorr}. A single peak structure
expected from PYTHIA calculations can not describe the measured
away side correlations. This observation of non-photonic e-h
correlation probably indicates heavy quark interaction with the
dense medium and the heavy quark energy loss may generate conical
emission in the hot dense matter created in heavy ion collisions
at RHIC.

As discussed above, the mechanism of the heavy quark interactions
with the dense medium is still not very clear. Thus it is of great
interest to separate non-photonic electron energy loss into the
contributions from charm and bottom quarks. Since the near-side
e-h azimuthal correlation from B decays is much wider than that
from D decays for the same electron $p_T$, STAR's previous
study~\cite{Lin07} compared the experimental correlation results
in p+p collisions with PYTHIA simulations, and found a substantial
B contribution to non-photonic electrons up to electron pt $\sim$6
GeV/$c$ (blue circles in the right panel of
Fig.~\ref{fig:ehcorr1}). And in Run7 this measurement has been
extended to pt $\sim$9 GeV/$c$ (blue triangles). Furthermore, STAR
presents another analysis technique to separate charm and bottom
quark contributions in the non-photonic electron measurement via
triggering on the leading non-photonic electron azimuthal
correlations with the balancing heavy quark identified by the
$D^0$ meson (Red circle and STAR)~\cite{AndreQM08}. The azimuthal
correlation distribution was studied using PYTHIA simulations and
simulations including NLO process~\cite{AndreQM08,NLO0203}, the
charm and bottom quark contribution can be separately estimated by
comparing the azimuthal correlation function from simulation and
data. PHENIX also presents similar analysis to separate charm and
bottom contribution in non-photonic electron measurement (purple
circles). A clear peak structure has been seen around 1.2
GeV/$c^2$ in the invariant mass distribution of triggered
non-photonic electron and correlated charged hadrons. By comparing
to the data, the difference of the invariant mass peaks of
triggered electrons and correlated charged hadrons from simulation
can be used to estimate the contributions from bottom and charm
quark for the same electron $p_T$~\cite{MorQM08}. All the
experimental results are consistent with the FONLL calculation
shown as the curves~\cite{cacciari}.

The upper limit of the FONLL calculation was used to estimate the
maximum contribution of electrons from bottom decays in the
non-photonic electron spectra, which were used to extract charm
cross section in a combined fit. In Fig.~\ref{cfitnob}, the solid
squares and solid circles are the non-photonic electron spectra in
0-12\% central and 0-80\% minbias Au+Au collisions at 200 GeV,
respectively. The open symbols present the non-photonic electron
spectra after the subtraction of the bottom contribution from the
upper limit of FONLL calculation. Then the bottom-excluded
non-photonic electron spectra were used in the new combined fit to
extract the charm production cross section. Table~\ref{crosssec}
shows the charm production cross sections before and after bottom
subtraction in minbias and central Au+Au collisions at 200 GeV.
The difference is within $\sim$5\%. Since the low $p_T$ muon
measurement sample most fraction of the charm cross
section~\cite{ffcharm}, the high $p_T$ bottom contribution does
not change it.

\begin{table}
\caption{\label{crosssec}Charm production cross sections before
and after bottom subtraction in minbias and central Au+Au
collisions at 200 GeV. The first error is statistical, the second
is systematical.} \centering
\begin{tabular}{|c|c|c|} \hline \hline
collisions & before bottom subtraction & after bottom
subtraction\\ \hline 0-12\% central Au+Au & $297\pm24\pm63\mu b$ &
$301\pm24\pm66\mu b$\\ \hline
0-80\% minbias Au+Au
&$274\pm25\pm60\mu b$ & $285\pm26\pm70\mu b$
\\ \hline \hline
\end{tabular}
\end{table}

\begin{figure}[htp]
\centering
\includegraphics[width=2.3in]{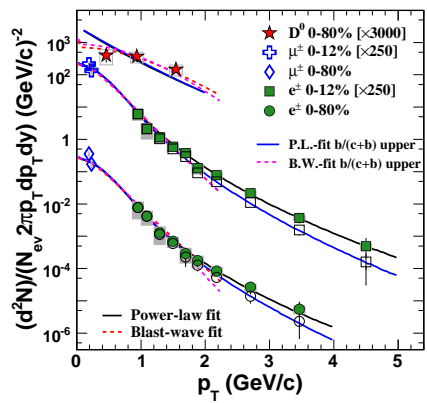}
\centering \caption{Combined fit to extract charm production cross
section using $D^0$, $\mu$ and non-photonic electron spectra
before (solid squares and circles) and after (open squares and
circles) bottom contribution
subtraction.}\vspace{-0.25cm}\label{fig:cfitnob}
\end{figure}

\section{Conclusions}
\vspace{-0.35cm}

The results of charm production from analysis of D meson
reconstruction and leptons from heavy flavor semileptonic decays
at RHIC were reported. The charm production cross section was
found to scale with number of binary collisions both in STAR and
PHENIX, which indicates that charm quark produced at early stage
of the system. But the discrepancy remains between STAR and
PHENIX. The blast-wave fits and the direct comparisons of the
spectra suggest that charmed hadrons interact with and decouple
from the system differently from the light hadrons. The non-zero
radial and elliptic flow observed at RHIC may suggest that the
light flavor thermalization at partonic stage in the hot dense
matter created in heavy ion collisions. The strong electron
suppression was observed both in STAR and PHENIX, the heavy quark
energy loss mechanism is not really understood. The broad
structure shown on the away-side e-h correlation may suggest the
heavy quark energy loss generates the conical emission in the
medium. The bottom contribution in the non-photonic electron
spectrum was studied by the e-h or e-D correlations, but the high
$p_T$ bottom contribution does not affect the charm production
cross section.

\end{document}